\documentclass[aip,cha,reprint,nofootinbib,a4paper]{revtex4-1}
\usepackage[compact]{titlesec}
\titleformat*{\section}{\large\bfseries}
\titlespacing{\section}{0pt}{8pt}{0pt}

\usepackage[utf8]{inputenc}
\usepackage{amsmath, amsthm, amssymb}
\usepackage{enumitem}
\usepackage{graphicx}
\usepackage{times}
\usepackage{dcolumn} 
\usepackage{url}
\usepackage{xcolor}
\usepackage{soul}
\DeclareMathAlphabet{\altmathcal}{OMS}{cmsy}{m}{n}

\usepackage[labelsep=space]{caption}

\DeclareCaptionFont{helv}{\fontfamily{helvet}\selectfont}

\begin{document}
\title{The role of susceptible individuals in spreading dynamics}

\author{Chang Su$^{1,3}$}
\author{Fang Zhou$^{1,3,*}$}
\author{Linyuan L\"u$^{2,1,*}$}

\footnotetext[1]{Institute of Fundamental and Frontier Sciences, University of Electronic Science and Technology of China, Chengdu 610054, P. R. China}
\footnotetext[2]{School of Cyber Science and Technology, University of Science and Technology of China, Hefei, 230026, P. R. China}
\footnotetext[3]{Yangtze Delta Region Institute (Huzhou), University of Electronic Science and Technology of China, Huzhou 313001, P. R. China}
\renewcommand{\thefootnote}{*}
\footnotetext{e-mail:zervel3@std.uestc.edu.cn; linyuan.lv@uestc.edu.cn}

\begin{abstract}

Exploring the internal mechanism of information spreading is critical for understanding and controlling the process. Traditional spreading models often assume individuals play the same role in the spreading process. In reality, however, individuals' diverse characteristics contribute differently to the spreading performance, leading to a heterogeneous infection rate across the system. To investigate network spreading dynamics under heterogeneous infection rates, we integrate two individual-level features---influence (i.e., the ability to influence neighbors) and susceptibility (i.e., the extent to be influenced by neighbors)---into the independent cascade model. Our findings reveal significant differences in spreading performance under heterogeneous and constant infection rates, with traditional structural centrality metrics proving more effective in the latter scenario. Additionally,  we take the constant and heterogeneous infection rates into a state-of-the-art maximization algorithm, the well-known TIM algorithm, and find the seeds selected by heterogeneous infection rates are more dispersed compared to those under constant rates. Lastly, we find that both individuals' influence and susceptibility are vital to the spreading performance. Strikingly, susceptible individuals are particularly important to spreading when information is disseminated by social celebrities. By integrating influence and susceptibility into the spreading model, we gain a more profound understanding of the underlying mechanisms driving information spreading.

\end{abstract}

\maketitle

\section*{Introduction}

In social networks, information spreads among individuals through specific dynamics. Understanding the role of individuals in the spreading process is significant to managing and controlling the entire network, which can be applied to areas such as virus marketing \cite{kim2006network, domingos2005mining, bhattacharya2019viral, arthur2009pricing}, rumor spreading \cite{borge2012absence, tong2018distributed, wang2021spreading} and political mobilization 
 \cite{cox1998mobilization, gonzalez2011dynamics, anduiza2014mobilization, pena2017individual}. 

To explore the information-spreading process, scholars have developed various spreading models. 
These models usually assume that individuals contribute equally to information spreading. One of the most popular and typical models is the Susceptible-Infected-Recovered (SIR) model \cite{kermack1927contribution}, which was initially proposed for disease spreading and later extended to social contexts. The SIR model assumes the same infection rates among contacted individuals. Such an assumption regards all individuals equivalently, rendering the SIR model a homogeneous spreading model with a constant infection rate. 
However, in reality,  individuals often possess different capacities to infect their neighbors, playing different roles in the spreading process. 
Thus, scholars put forward some heterogeneous spreading models, such as the linear threshold model and the independent cascade model \cite{chen2010scalable, saito2008prediction}, where the infection rates of edges can be different. Specifically, the linear threshold model assumes that an inactive individual becomes activated when the sum of its infected neighbors' weight exceeds a threshold value, while the independent cascade model assumes that different edges hold different spreading probabilities, with an infected individual influencing its neighbors at a corresponding infection rate.

While these heterogeneous models have taken into account the varied roles of individuals in the network,  they have primarily focused on how structure-based features \cite{kitsak2010identification, chen2012identifying, lu2016vital, morone2015influence} affect the spreading dynamics, neglecting consideration of individual-level features. 
Recently, Sinan Aral et. al considered two features for individuals: influence, i.e., the ability to influence neighbors, and susceptibility, i.e., the extent to which one can be influenced by neighbors\cite{aral2012identifying}. 
They proposed that the infection rate between two individuals is a combination of the source individual's influence and the target individual's susceptibility. 
Their large-scale empirical social experiments revealed that both influential and susceptible individuals can exert a significant impact on spreading dynamics. 
In his later work\cite{Sinan2018Social}, Aral further demonstrated that, in comparison to the spreading model incorporating individuals' influence and susceptibility, social impact under the traditional homogeneous spreading model has been greatly underestimated \cite{Sinan2018Social}.


However, in the paper \cite{Sinan2018Social},  Aral generated empirical models according to the assortativity of the joint distribution of influence and susceptibility, without separately exploring the effects of individuals' influence and susceptibility.
Thus, how these two features drive the spreading process remains unknown, and the ongoing debate is yet to be resolved. Supporters of the "influential hypothesis"  assert that influential individuals are the primary drivers of the diffusion of information, behaviors, and markets in society \cite{aral2012identifying}, while supporters of the "susceptibility hypothesis" argue that susceptibility plays the key role in driving social contagion \cite{watts2007influentials, dodds2004universal, centola2007complex}. 
To solve this controversy, 
we design a series of experiments to examine the roles of influential and susceptible individuals in the spreading process.

Our paper is structured as follows: Firstly, we introduce the methods, focusing on the independent cascade model,  traditional centrality metrics, and the datasets used in our experiments. 
Then,  we compare spreading dynamics under constant and heterogeneous infection rates. 
Furthermore, we delve into the scenario of heterogeneous infection rates and investigate the roles of influence and susceptibility in both normal-individuals-driven and celebrities-driven spreading patterns.
Finally, we summarize our findings and discuss future research.

\section*{Methods}
\subsection*{Independent cascade model}


In this paper, we employ the widely-used heterogenous spreading model---the independent cascade (IC) model \cite{Sinan2018Social} to simulate information spreading on social networks. Specifically, given a network $G$ and edges with infection rate $p_{ij}$, we choose an individual $i$ or a group of individuals as seeds to initiate the spreading process. Once an individual is infected, it has only one opportunity to infect each of its neighbors.  The spreading process continues until there are no more individuals to infect, and the final size of infected individuals is referred to as the spreading capacity of individual $i$.

Within this model, we assume that each individual has two features: influence and susceptibility. 
Different from the traditional definition of influence as a global metric affecting individuals of the whole network, here, an individual's influence refers to its ability to affect its nearest neighbors. Similarly, susceptibility is the extent to which an individual is affected by its neighbors. 
The infection rate between two linked individuals $i$ and $j$ is denoted as $p_{ij}=I_i S_j$, where $I_i$ is the influence score of individual $i$, and $S_j$ is the susceptibility score of individual $j$.

\subsection*{Centrality metrics}
We employ two types of centrality metrics in the experiments: unweighted-based metrics and weighted-based metrics. 
\paragraph{Degree}
For a given individual $i$, its degree is the number of nearest neighbors \cite{newman2018networks}, which is 
\begin{equation}
k_i = \sum_{j=1}^N A_{ij}.
\end{equation}
Here,  $A$ refers to the corresponding adjacent matrix of network $G$. $A_{ij}$=1 if individuals $i$ and $j$ are linked; otherwise, $A_{ij}$=0.


\paragraph{k-core}
In an unweighted graph, for an individual $i$, its k-core score is $k$ if and only if node $i$ belongs to a maximal
subgraph whose members have a degree of at least $k$ \cite{kitsak2010identification}.

\paragraph{H-index}
The H-index was first introduced to measure the academic impact of scholars and was later adapted for use in complex networks to measure the influence of nodes \cite{lu2016h}. For node $i$ with degree $k_i$ and for each of its neighbors with degree $k_{j_1}$, $k_{j_2}$,..., $k_{j_{k_i}}$, the H-index of node $i$ is $h$ when $i$ has at least $h$ neighbors whose degrees are at least $h$. In the formula, it can be expressed as

\begin{equation}
    h_i = H(k_{j_1}, k_{j_2},...,k_{j_{k_i}}).
\end{equation}

\paragraph{Eccentricity}
The eccentricity score of node $i$ is defined as the maximum distance among all the shortest paths to the other nodes \cite{hage1995eccentricity}. In the formula, it can be expressed as

\begin{equation}
    ECC(i) = \max_{{v_j}\neq{v_i}} \{d_{ij}\} .
\end{equation}

Here, $d_{ij}$ is the shortest path length between node $i$ and node $j$.

\paragraph{Closeness}
The closeness score of node $i$ is defined as the reciprocal of the average shortest path distance from node $i$ to other nodes \cite{freeman1978centrality}.   The average shortest path length between node $i$ and other $N-1$ nodes can be expressed as

\begin{equation}
    d_i = \frac{1}{N-1} \sum_1^{N} d_{ij} .
\end{equation}


The closeness score $CC_i$ of node $i$ is
\begin{equation}
    CC_i = d_i^{-1}.
\end{equation}

\paragraph{Betweenness}
The betweenness score of node $i$ is defined as the fraction of the shortest paths that pass through node $i$ 
 \cite{brandes2001faster}. In the formula, it can be expressed as

\begin{equation}
    BC_i = \sum_{s\neq t \neq i}\frac{n_{st}^i}{g_{st}}.
\end{equation}

Here, $n_{st}^i$ is the number of the shortest paths from node $s$ to $t$ that go through node $i$, and $g_{st}$ is the number of the total shortest paths.

\paragraph{PageRank}
The PageRank algorithm was initially developed to rank websites in the Google search engine and was subsequently applied to other commercial scenarios \cite{haveliwala2002topic}. 
This algorithm works as follows: 
First, each node is assigned a PR value of one unit. 
Then every node evenly distributes its PR value to its neighbors along its outgoing links. Mathematically, the PR value of node $v_i$ at $t$ step is
\begin{equation}
    PR_i^{(t)} = \sum_{j=1}^{n} a_{ij}\frac{PR_j ^{(t-1)}}{k_{j}^{out}},
\end{equation}
where $n$ is the total number of nodes in the network, and $k_j^{out}$ is the out-degree of node $v_j$.
The iteration will stop if the PR values of all nodes reach a steady state.



\paragraph{Weighted Degree}
For a given individual $i$, its weighted degree is the sum of its nearest edges' weight \cite{opsahl2010node}, which is 
\begin{equation}
WD_i = \sum_{j=1}^N p_{ij}.
\end{equation}

Here,  $p_{ij}$ refers to the infection rate from node $i$ to node $j$.

\paragraph{Weighted H-index}
The weighted H-index score of node $i$ is calculated by the $\mathcal{H}$ function, which operates on the series of out-strength values of $v_i$’s neighbors \cite{wu2019coreness}. The $\mathcal{H}$ function would return the maximum real number $x$ which satisfies $f(x) \ge x$, where
\begin{equation}
    f(x) = \left\{ 
    \begin{array}{lr}
    s_{j_1}\ if\ 0< x \leq w_{ij_1} \\
    s_{j_r}\ if \sum_{m=1}^{r-1}w_{ij_m}< x \leq \sum_{m=1}^{r}w_{ij_m} \ for\ r\ge 2.
    \end{array}
    \right.
\end{equation}
Compared with the H-index, the weighted H-index takes extra consideration for path weighted. In the formula, it can be expressed as
\begin{equation}
    h_i^W = \mathcal{H}[(w_{{ij_1}},{s_{j_1}}),(w_{{ij_2}},{s_{j_2}}),...,(w_{{ij_k}},{s_{j_k}})].
\end{equation}


\paragraph{Weighted Closesness}
In a weighted network, the first step is to redefine weighted-shortest paths \cite{opsahl2010node}. In the formula, it can be expressed as
\begin{equation}
    d_{ij} = \min\left( w_{ih_0}+w_{h_0 h_1}+...+w_{h_k j} \right),
\end{equation}
where $v_{h0}, v_{h1},..., v_{hk}$ are the intermediary nodes belonging to a path from $v_i$ to $v_j$.
Then, the weighted closeness score of node $i$ is defined as
\begin{equation}
    WCC_i = \left[ \sum_j^n d_{ij}^w \right]^{-1}.
\end{equation}

\paragraph{Weighted Betweenness}
The weighted betweenness score of node $i$ is defined as \cite{opsahl2010node}

\begin{equation}
    {WBC}_i = \sum_{s\neq t \neq i}\frac{g_{st}^w(i)}{g_{st}^w},
\end{equation}
where $g_{st}^w$ is the number of the total shortest paths from $v_s$ to $v_t$, and $g_{st}^w(i)$ is the number of the shortest paths from $v_s$ to $v_t$ that pass through node $v_i$.

\paragraph{Weighted PageRank}
In the weighted PageRank, the PR value of a node will be distributed to its outgoing neighbors according to the link weights \cite{xing2004weighted}. In the formula, it can be expressed as

\begin{equation}
    WPR_i^{(t)} = \sum_{j=1}^{n} w_{ij}\frac{WPR_j ^{(t-1)}}{s_{j}^{out}},
\end{equation}
where $s_{j}^{out}$ is the out-strength of $v_j$.

\subsection*{Datasets}
For each simulation, we conduct it in six datasets:  five real-world social datasets (Arenas-Email, Facebook, HepTh, Hamster, and LastFM-Asia), and SWP10, a small-world synthetic dataset with a $10\%$ connection probability. 
Specifically, Arenas-Email is a communication network of a university in Spain \cite{guimera2003self}; nodes represent users, and an edge exists between two users if at least one email was sent between them.
Facebook is a friendship network sourced from the social platform \emph{Facebook}  \cite{snapnets}; nodes represent users, and an edge exists between two users if they are friends.
HepTh is a collaboration network at the open-access publication platform \emph{arXiv}  \cite{snapnets}; nodes represent authors, and an edge exists between two authors if they have collaborated on at least one publication.
Hamster is a friendship network from the social media platform \emph{Hamsterster} \cite{kunegis2013konect}; nodes represent users, and an edge exists between two users if they are friends.
LastFM-Asia is a social network of \emph{LastFM} users \cite{rozemberczki2020characteristic}; nodes represent users, and an edge exists between two users if they follow each other on the platform.

Table \ref{table1} shows the structural details of the six datasets. $N$ is the size of the network. $L$ is the sum of edges. $\langle k \rangle $ is the average degree. $\sigma$ is the degree assortativity. $c$ is the clustering coefficient.

\begin{table}[!ht]
\centering
\caption{Structural details of the six datasets.}
\begin{tabular}{ccccccc}
	\hline
	datasets  & $N$ \  & $L$ &  $\langle k \rangle $ &$\sigma$ & $c$ \\
	\hline

	Arenas-Email & 1,133 & 5,451 & 9.622 & 0.078 & 0.166 \\

    Facebook & 4,039 & 88,234 & 43.691 & 0.064 & 0.519\\

	HepTh & 9,875 & 25,973 & 5.260 & 0.267 & 0.283\\

	Hamster & 1,858 & 12,534 & 13.491 & -0.847 & 0.090\\

	LastFM-Asia & 7,624 & 27,806 & 7.294 & 0.0171 & 0.179\\	
    SWP10 & 5,000 & 20000 & 8 & 0.011 & 0.161 \\

	\hline
\end{tabular}
\label{table1}
\end{table}

\section*{Results}
\subsection*{Spreading dynamics under constant and heterogeneous infection rates}

To understand the different spreading dynamics of networks under a constant infection rate and a heterogeneous one, we employ the IC model to simulate the spreading process and calculate the correlation between individuals' spreading capacity (the fraction of infected individuals) and their  $12$ centrality metrics scores.

Specifically, we develop two types of IC models: one with a constant infection rate and another with a heterogeneous infection rate. Within each type, there are two distinct models. 
For two constant infection rate models, we set the infection rate as $p_{ij}=avg\left\{S_{j}\right\}$ and $p_{ij}=constant$, respectively, where $avg\left\{S_{j}\right\}$ is the averaged susceptibility score of $i$'s neighbors, and $constant$ is fixed at $0.1$.  
For two models with heterogeneous infection rates, we set the infection rate as $p_{ij}=I_{i}$ and $p_{ij}=I_{i} S_{j}$, respectively, where $I_{i}$ is the influence score of individual $i$, and $S_j$ is the susceptibility score of individual $j$.

For each model, the spreading process operates as follows. 
First, for each individual in network $G$, we generate its influence and susceptibility scores by uniform distribution from $0$ to $1$. 
Then, we calculate the infection rate of each edge based on the defined infection rates mentioned above. Next, each individual is selected one at a time as the seed for information spreading. This process is repeated $100$ times, and the spreading capacity of each individual is determined by averaging the results from these iterations. 
Finally, we use four coefficients ---Pearson, Spearman, Kendall, and "top $10\%$ precision---to measure the relationship between spreading capacity and the seed's $12$ centrality scores (see Methods). Pearson, Spearman, and Kendall are all defined as the correlation between individuals' spreading capacity and their centrality scores, while "top $10\%$ precision" is defined as the fraction of the overlap between individuals ranked in the top $10\%$ based on centrality score and those ranked in the top $10\%$ based on spreading capacity.
For each model, the final correlation result is determined by averaging across $12$ centrality metrics and six datasets.

As shown in table \ref{table0}, the correlation between nodes' spreading capacity and metrics scores is much higher under constant infection rates  ($p_{ij}=avg\left\{S_{j}\right\}$, $p_{ij}=constant$) than under heterogeneous infection rates ($p_{ij}=I_{i}$, $p_{ij}=I_{i} S_{j}$). 
This indicates that when the infection rate is constant, the structure centrality metrics are more effective in identifying individuals who are important for spreading information. However, when the infection rate is heterogeneous, these metrics are not as useful in identifying important individuals. 
For detailed correlation and precision values between individuals' centrality scores and spreading capacity under different infection rates, please refer to Tables A1, A2, and A3 in Appendix A.

\begin{table}
\caption{{\bf The correlation between individuals' spreading capacity and centrality metrics scores under constant and heterogeneous infection rates.} The first two rows ($p_{ij}=constant$ and $p_{ij}=avg\left\{S_{j}\right\}$ ) show the results under a constant infection rate, while the last two rows  ($p_{ij}=I_{i}$ and $p_{ij}=I_{i} S_{j}$) show that under a heterogeneous rate. }

\scalebox{1}{
\begin{tabular}{ccccc}
\hline

  & Pearson & Spearman &  Kendall & top10\% precision \\
	\hline
    $constant$ & $0.86$ & $0.91$ & $0.77$ & $0.70$  \\
    $avg\left\{S_{j}\right\}$ & $0.42$ & $0.56$ & $0.43$ & $0.36$ \\
    $I_i$ & $0.35$ & $0.50$ & $0.37$ & $0.24$  \\
    $I_i S_j$ & $0.42$ & $0.55$ & $0.42$ & $0.35$\\
	\hline
\end{tabular}

}

\label{table0}
\end{table}

Then, how does the selection of key individuals, or the seeding policy, differ between models under constant infection rates and heterogeneous ones? 
To solve this, we convert the seeding policy problem into an influence maximization problem, whose goal is to find a subset of individuals in a network to maximize the spreading throughout the network. 
As the optimal solution to the influence maximization problem is computationally difficult, here, we employ a heuristics algorithm---the TIM algorithm \cite{tang2014influence}---to find the optimal seeds. 
Specifically, in a round, we generate individuals' influence and susceptibility scores. Then, we take either a constant infection rate $p_{ij}=constant$ or a heterogeneous infection rate $p_{ij}=I_i S_j$ as input to find the optimal seeds. Finally, we record the degree of selected seeds.  This process is repeated $10$ times.

Fig. \ref{fig:timseeddegree} shows the degree distribution of seeds selected under constant and heterogeneous infection rates across six datasets.
It can be found that the degree of optimal seeds selected under a heterogeneous infection rate is more dispersed than seeds selected under a constant infection rate. 
This suggests that the inclusion of heterogeneous infection rates can broaden the selection of seeds, and not just limit it to individuals that are highly central in the network.
Thus, in practical scenarios with varying infection rates, such as in social marketing, it is advisable for companies not only to rely on socially central celebrities as seeds but also to consider a more diverse range of individuals. 


\begin{figure*}[!htb]
    \centering

    \caption{ Degree distribution of seeds selected under two types of infection rates. The $x$-axis represents the infection rate types: a constant infection rate (denoted as "constant"), and a heterogeneous infection rate (denoted as 
 "i$\times$s"). The $y$-axis represents the degree distribution of the selected seeds. 
 In each realization, the seeds are represented as scatter dots in a column, and a total of $10$ realizations are recorded.  }

    \label{fig:timseeddegree}
\end{figure*}

\subsection*{Roles of influential and susceptible individuals in spreading dynamics}

In the above section, we reveal that a heterogeneous infection rate can affect the spreading performance and optimal seed selection. To comprehend this impact, we narrow our focus to two features influencing the spreading dynamics: influence and susceptibility. 
Specifically, we investigate the roles of these two features in spreading from two aspects: when individuals serve as seeds and when they do not. 

 \subsubsection*{Influential and susceptible individuals as seeds}

\begin{figure*}[!htb]
    \centering
    \includegraphics[scale=0.2]{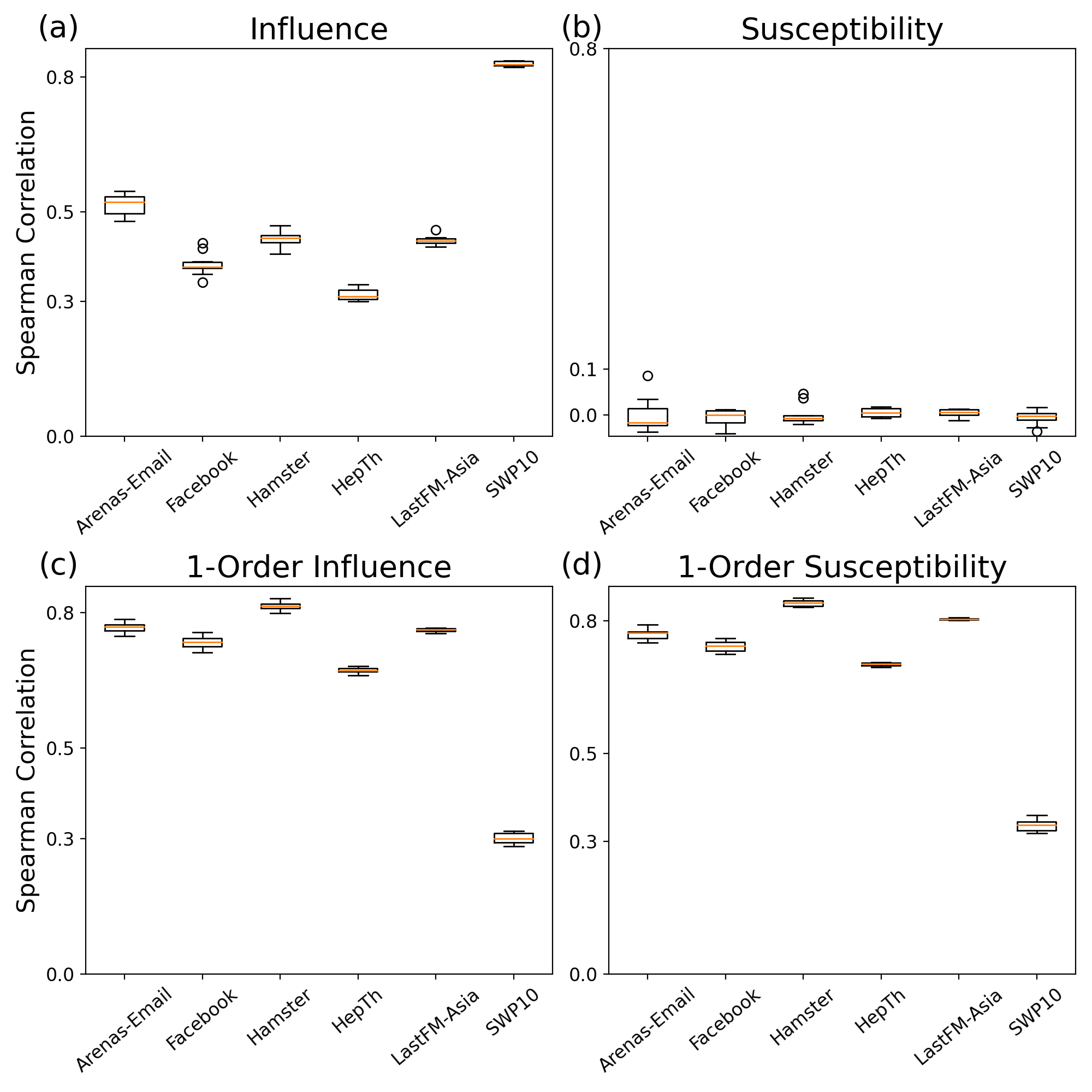}
    \caption{ {\bf The Spearman correlations between individuals' spreading capacity and two features: influence and susceptibility. }
Panels  (a) and (b) display the correlation between seeds' spreading capacity and their influence and susceptibility, respectively. Panels (c) and (d) depict the correlation between seeds' spreading capacity and their $1$-order influence and 1-order susceptibility respectively.} 
    \label{fig:fig1}
\end{figure*}

In this experiment, we take individuals with different influence and susceptibility scores as seeds to start the spreading process and analyze which type of individuals are more crucial. 
First, for a given network $G$, we generate individuals' influence and susceptibility scores randomly. Then, we choose each individual one at a time as the seed to start the IC spreading process; this process is repeated $1000$ times for each seed. Finally, we calculate the Spearman correlation between the individuals' spreading capacity (averaged over $1000$ realizations) and their corresponding influence and susceptibility scores. The results are presented in Fig. \ref{fig:fig1} (a) and (b). We find that the seed's influence score is correlated with the final spreading capacity, with the average corresponding Spearman coefficients ranging from $0.3$ to $0.859$ across the six datasets. This indicates that a seed with a high influence score tends to infect more individuals in the spreading process. However, a seed's susceptibility score shows little correlation with the final spreading size, with the Spearman coefficient nearly approaching  $0$ across the six datasets. This reveals that the seed's susceptibility can hardly affect its final spreading capacity under the above assumption. 

To analyze why seed susceptibility is almost uncorrelated with the final spreading capacity, we turn back to the IC spreading process, where seed $i$ initiates the spreading process and infects its neighbor $j$ with probability $p_{ij}=I_i S_j$. We observe that the susceptibility score $S_i$ of seed $i$ does not participate in the subsequent spreading process, thus showing little influence on the spreading capacity. This means that using this approach to assess the impact of influence and susceptibility on spreading capacity is not appropriate and we need to compare these two features under comparable conditions.

Therefore, for a seed individual $i$, we turn to its $1$-order neighbors' influence and susceptibility. Specifically, we define two $1$-order metrics: the sum of $i$'s neighbors' influence score ($\sum_j A_{ij}I_j$), and the sum of $i$'s neighbors' susceptibility score ($\sum_j A_{ij}S_j$). Then, we choose each individual one at a time as the seed to start the spreading process; this process is repeated $1000$ times for each seed. Finally, we calculate the Spearman correlation between spreading capacity (averaged over $1000$ realizations)  and their corresponding 1-order influence and 1-order susceptibility scores. 
The results are displayed in Fig.\ref{fig:fig1} (c) and (d).  It can be found that the Spearman correlations for the two metrics are similar across the six datasets, with the coefficients being more than $0.7$ in five of the six networks. 
This indicates that both individuals' influence and susceptibility play important roles in the spreading process when they are seed nodes. Neglecting either of them will hinder the spreading of information. We also calculate the Pearson and Kendall correlations between the spreading capacity of seeds and the two metrics, and the results remain consistent. See Appendix B Figs. A1 and A2 for details.
 
\subsubsection*{Influential and susceptible individuals as normal nodes}

The above results revealed that if we take individuals with high $1$-order influence and susceptibility as seeds, they can achieve almost the same final spreading effect, indicating both two features play important roles in the spreading process. If highly influential and susceptible individuals are not seeds, how will their deficiency affect the final spreading effect?
To comprehend the impact of influential and susceptible nodes on spreading dynamics when they function as normal nodes rather than seed nodes,  we remove these nodes from the network and compare the resulting spreading size. For comprehensive experiments,  we employ two types of spreading patterns: one driven by normal individuals (randomly selected nodes as seeds), and another driven by celebrities (high-degree nodes as seeds).

Specifically, for the normal-individuals-driven pattern, we first select $1\%$  of nodes as seeds randomly. 
Then, we remove individuals using one of the three strategies with a fraction ranging from $0\%$ to $50\%$: (\romannumeral1) removing individuals with high influence score; (\romannumeral2) removing individuals with high susceptibility score; (\romannumeral3) removing individuals randomly. Finally, we use the IC model to simulate the spreading process and the results are averaged over $100$ realizations.  We also explore the use of $100$ nodes and $0.5\%$ of nodes as seeds and see Appendix C Figs. A5 and A6 for details.

\begin{figure*}[!htb]
    \centering
    \includegraphics[scale=0.3]{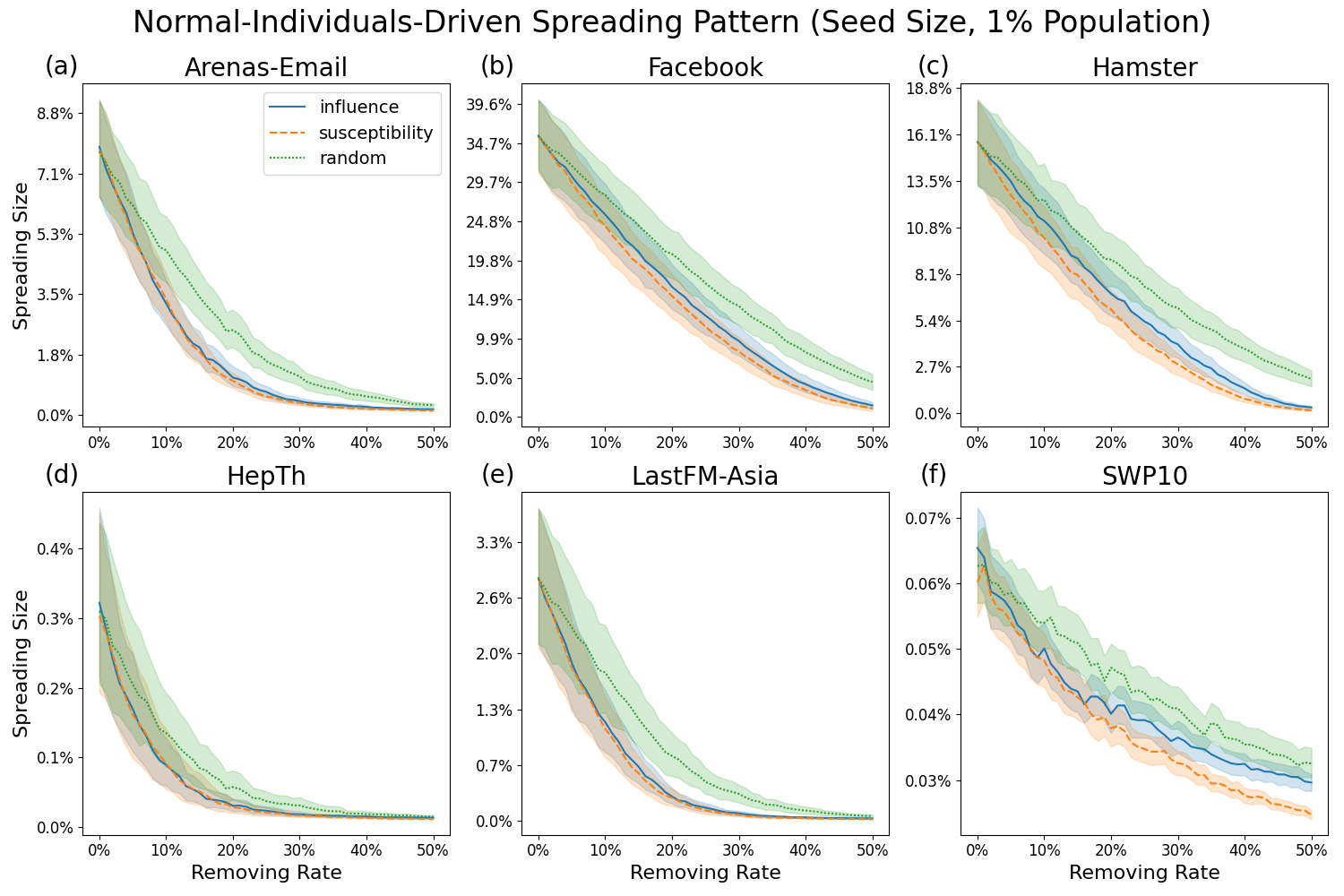}
    \caption{{\bf The impact of individuals' influence and susceptibility on spreading under the normal-individuals-driven spreading pattern.} Under this pattern, we select $1\%$ of nodes as seeds randomly. 
    The $x$-axis represents the fraction of individuals removed, while the $y$-axis represents the spreading capacity of the seeds. 
    The blue, red, and green curves represent three strategies of node removal:  removing nodes by influence, by susceptibility, and removing nodes randomly, respectively; the shadow around each curve denotes the spreading capacity with a $95\%$ confidence interval.}
    \label{fig:normal}
\end{figure*}

The results are displayed in Fig. \ref{fig:normal}. We find that, except for the SWP10 dataset, all three strategies lead to an obvious decline (up to $35\%$) in the network's spreading size. Notably, strategy 1 (removing influential individuals) and strategy 2 (removing susceptible individuals) demonstrate a similar degree of impact on spreading size, outperforming strategy 3 (removing individuals randomly). 
Regarding the SWP10 dataset, all three strategies yield almost negligible decline (less than $0.1\%$ ) in spreading capacity, leading us to infer potential difficulties in information diffusion for seeds within this network. 
To mitigate the influence of poor diffusion, we additionally employ $3$ synthetic datasets with a relatively large average degree. The results are consistent with those of the other five datasets, confirming the comparable impact of targeting influential and susceptible nodes on spreading dynamics in networks driven by normal individuals.  Please refer to Appendix C Fig. A9 for details.

After the experiment under the normal-individuals-driven pattern, we extend our investigation to the celebrities-driven pattern, where seeds are selected by degree, not randomly. 
Here, we select the top $1\%$ of individuals with the highest degree as seeds (see Appendix C Figs. A3 and A4 for details on alternative seed selection fractions), and remove individuals using the same three strategies with a fraction ranging from $0\%$ to $50\%$;  the simulation results are also averaged over $100$ realizations. 

\begin{figure*}[!htb]
    \centering
    \includegraphics[scale=0.25]{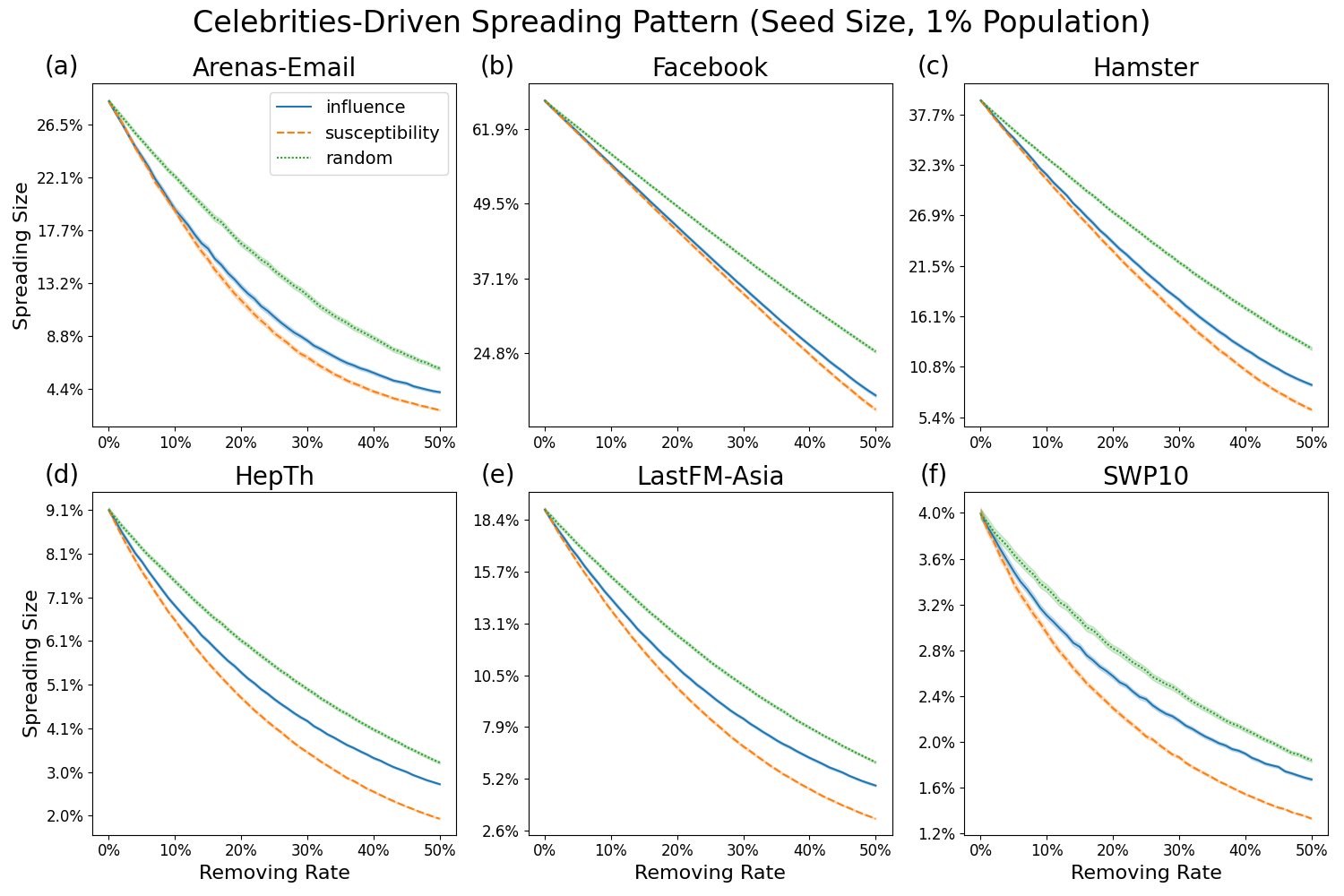}
    \caption{ {\bf The impact of individuals' influence and susceptibility on spreading under celebrities-driven spreading pattern.} 
Under this pattern, we select the top $1\%$ of nodes as seeds by degree.  } 
    \label{fig:celebrity}
\end{figure*}

The results are depicted in Fig. \ref{fig:celebrity}, yielding two conclusions. 
First, consistent with the preceding experiment, strategies $1$  and $2$ lead to a greater decline in spreading capacity than strategy $3$, suggesting that influential or susceptible individuals play a more important role than common individuals in transmitting information.
Second, in all $6$ datasets,  removing susceptible individuals leads to a greater decline in spreading size than removing influential individuals. This indicates that, under the celebrities-driven pattern, the removal of highly susceptible individuals has a deeper impact on spreading dynamics than the absence of influential individuals. Thus, under this condition, highly susceptible individuals are deemed more crucial than highly influential ones.

\begin{figure*}[!htb]
    \centering
    \includegraphics[scale=0.25]{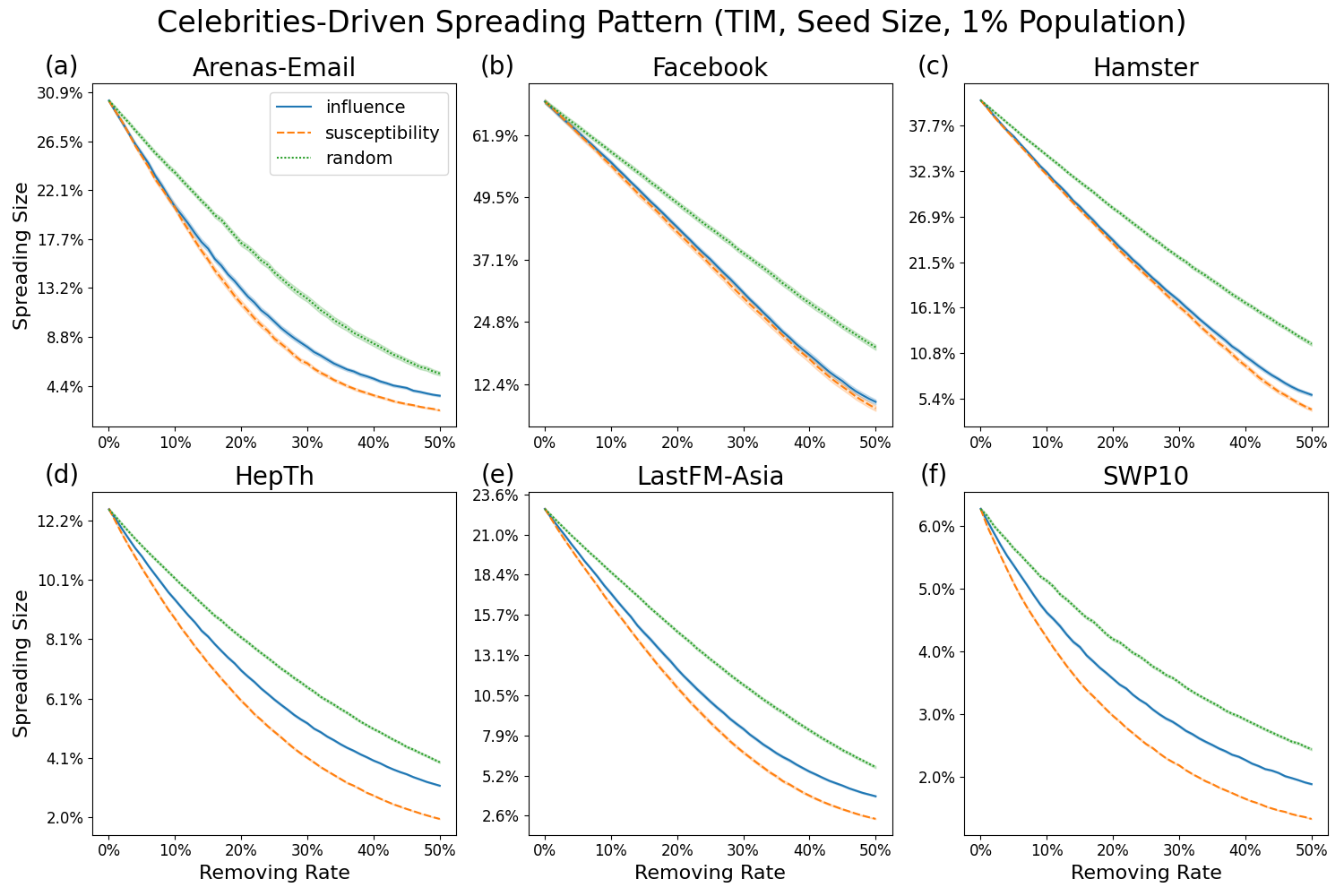}
    \caption{{\bf The impact of individuals' influence and susceptibility on spreading under celebrities-driven spreading pattern.} Here, we use the TIM algorithm to select $1\%$ of nodes as seeds, instead of choosing seeds based on the degree.}
    \label{fig:TIM}
\end{figure*}

The insight into the more crucial role of highly susceptible individuals in the celebrities-driven spreading pattern is valuable. However, this observation could also be attributed to the overlap in the structural position of celebrities, known as the rich-club effect \cite{senden2014rich}. 
To test whether node position or its susceptibility feature leads to its greater impact, we employ the TIM algorithm to choose a different set of seeding nodes and conduct the spreading experiment.
The results, illustrated in Fig. \ref{fig:TIM},  align with the preceding ones: the removal of nodes with high susceptibility scores leads to the most substantial decline in network spreading capacity. 
This confirms that the structural position is not the factor leading to the performance difference between strategies $1$ and $2$, and highly susceptible individuals indeed play a more significant role in celebrities-driven spreading process compared to influential ones.
We also conduct tests with the top $100$ and $0.5\%$  of seeds selected by the TIM algorithm, and the results are consistent. Please see Appendix C Figs. A7 and A8 for details.

\section*{Conclusion and discussion}
In this paper, we integrate two features of individuals, namely influence and susceptibility, into the IC spreading model to discuss their roles in spreading dynamics. 
We first explore the spreading dynamics under constant and heterogeneous infection rates. Our findings demonstrate that the spreading performance of individuals under two types of infection rates differs significantly from that under a constant infection rate, and structural centrality metrics are more suitable for identifying the spreading capacity of individuals when the dynamics are activated by constant infection rates. Additionally, there are distinguishable patterns in seeding policies under heterogeneous and constant infection rates. Specifically, the degrees of seeds chosen under a heterogeneous infection rate are more dispersed, while those chosen under a constant infection rate exhibit a more centralized distribution.


Then, we delve into the roles of influential and susceptible individuals in the spreading process, considering two perspectives: when they serve as seeds and when they do not.
We observe that both influential and susceptible individuals play significant roles in the spreading process when acting as seeds. There exists a strong correlation between individuals' $1$-order influence and susceptibility scores and their spreading capacity, indicating that individuals with these features have the potential to infect a larger number of individuals, leading to large-scale spreading. 
Nonetheless, when influential and susceptible individuals do not function as seeds in the network, their roles can differ. Specifically, when the network is driven by normal individuals, the removal of highly influential and susceptible individuals yields nearly identical impacts. 
However, when the spreading process is driven by high-degree celebrities (hub individuals in networks), the shortage of highly susceptible individuals results in a more substantial decline in spreading capacity. This outcome underscores that susceptible individuals play a greater role than influential ones in spreading when information spreads starting from powerful celebrities.

 To date,  most studies have focused on how individuals' structure-based metrics affect the spreading performance \cite{kitsak2010identification, chen2012identifying,lu2016vital, morone2015influence}.  Through field experiments and simulation, the authors have discovered that influential and susceptible individuals are crucial in the spreading process \cite{aral2012identifying, Sinan2018Social}. In general, there have been few studies that discuss how influential and susceptible individuals affect information spreading simultaneously.
Previous studies have gotten different conclusions about which type of individual has a greater impact on the spreading process.  The influential hypothesis suggests that influential individuals catalyze the diffusion of information, behaviors, innovations, and products in society \cite{katz1957two,valente1995network, kim2009analytical, aral2012identifying, roelens2016identifying}, which is the mainstream conclusion accepted by researchers. Conversely, other studies find that susceptibility drives the spreading of information \cite{ dodds2004universal,  watts2007influentials, aral2012identifying}. The above works involved influential and susceptible individuals from different perspectives. Different from the paper \cite{Sinan2018Social} considers the assortativity of the joint distribution of influence and susceptibility to influence maximization problem, which is from a macroscopic perspective. Our study takes a microscopic perspective and aims to resolve this controversy theoretically by incorporating both individuals' influence and susceptibility into the independent cascade model. Finding both influential and susceptible individuals plays an important role in the spreading process, and under the celebrities-driven pattern, the susceptible individual is more crucial in the spreading process. 

Considering that the spreading of information is usually driven by celebrities in social networks \cite{ho2016social, jin2018celebrity, zhou2020realistic}, our finding about the crucial role of susceptible individuals in celebrities-driven networks provide valuable practical implications. 
It can be applied in various information strategy designs, from controlling the spreading of rumors to accelerating the propagation of a message.
To substantiate this conclusion, additional real-world experiments and research are needed. This could potentially become a focus of our future studies, and we also hope our work serves as an inspiration for researchers to delve deeper into this area.

\bibliography{references}

\section*{Acknowledgment}
The authors acknowledge support from the National Natural Science Foundation of China (Grant No. T2293771), the STI 2030—Major Projects (Grant No. 2022ZD0211400), the Sichuan Province Outstanding Young Scientists Foundation (Grant No. 2023NSFSC1919), and the New Cornerstone Science Foundation through the XPLORER PRIZE. 
\end{document}